\PassOptionsToPackage{table}{xcolor}
\documentclass[acmsmall,screen]{acmart}
\pdfoutput=1
\usepackage{xspace}
\usepackage{xcolor}
\usepackage{amsmath}
\usepackage{algorithmic}
\usepackage{array}
\usepackage[caption=false,font=footnotesize]{subfig}
 \usepackage{grffile}
\usepackage{url}
\usepackage{lipsum}
\usepackage{textcomp}
\usepackage{tcolorbox}
\usepackage{multirow}

\newcommand{\eg}{\hbox{\emph{e.g.,}}\xspace}
\newcommand{\ie}{\hbox{\emph{i.e.,}}\xspace}

\newcommand{\prompt}[4]{
    \begin{figure}[htpb]
    \vspace{-0.1in}
        \centering
        \begin{tcolorbox}[fontupper=\footnotesize,
            colframe=black, colback=white, coltitle=white, colbacktitle=black,
            title=#1 %, label=#3
        ]
            \textcolor{black}{
                #2
            }
        \end{tcolorbox}
        \vspace{-0.1in}
        \caption{#4}
         \label{#3}
    
    \end{figure}
    \vspace{-0.2in}
}

\usepackage{cleveref}

\Crefname{figure}{Figure}{Figures}
\crefname{figure}{Figure}{Figures}

\crefname{listing}{Listing}{Listings}
\Crefname{listing}{Listing}{Listings}

\Crefname{table}{Table}{Tables}
\crefname{table}{Table}{Tables}

\Crefname{box}{Figure}{Figures}
\crefname{box}{Figure}{Figures}

\crefformat{chapter}{\S~#2#1#3}
\crefmultiformat{chapter}{\S\S~#2#1#3}{ and~#2#1#3}{, #2#1#3}{, and~#2#1#3}

\crefformat{section}{Section~#2#1#3}
\crefmultiformat{section}{\S\S~#2#1#3}{ and~#2#1#3}{, #2#1#3}{, and~#2#1#3}

\usepackage{hyperref}

\newcommand{\rqone}{\textit{RQ1: How effective is \ourTool compared with SOTA techniques?}\xspace}

\newcommand{\rqtwo}{\textit{RQ2: How effective is each component in \ourTool?}\xspace}
    
\newcommand{\rqthree}{\textit{RQ3: Can the analysis generated by \ourTool help security experts in identifying vulnerability fixes?}\xspace}

\newcommand{\rqfour}{\textit{RQ4: Bad Case Analysis: In which scenarios does \ourTool fail?}\xspace}

\newcommand{\ourTool}{\mbox{LLM4VFD}\xspace}

\newcommand{\rqboxc}[1]{\begin{tcolorbox}[left=0.5pt,right=0.5pt,top=0pt,bottom=0pt,colback=gray!5,colframe=gray!40!black,before skip=1pt,after skip=0pt]#1\end{tcolorbox}}

\usepackage{soul}

\begin{document}

\title{Code Change Intention, Development Artifact and History Vulnerability: Putting Them Together for Vulnerability Fix Detection by LLM}

\author{Xu Yang}
\email{yangx4@myumanitoba.ca}
\affiliation{%
  \institution{University of Manitoba}
  \city{Winnipeg}
  \state{Manitoba}
  \country{Canada}
}
\author{Wenhan Zhu}
\email{wenhanzhu1@acm.org}
\affiliation{%
  \institution{Huawei Canada}
  \country{Canada}
}
\author{Michael Pacheco}
\email{michael.pacheco1@huawei.com}
\affiliation{%
  \institution{Huawei Canada}
  \country{Canada}
}

\author{Jiayuan Zhou}
\email{jiayuanzhou1@acm.org}
\affiliation{%
  \institution{Huawei Canada}
  % \city{Winnipeg}
  % \state{Manitoba}
  \country{Canada}
}

\author{Shaowei Wang}
\authornote{Corresponding author.}
\email{shaowei.wang@umanitoba.ca}
\affiliation{%
  \institution{University of Manitoba}
  \city{Winnipeg}
  \state{Manitoba}
  \country{Canada}
}

\author{Xing Hu}
\email{xinghu@zju.edu.cn}
\affiliation{%
  \institution{Zhejiang University}
  \city{Hangzhou}
  \country{China}
}

\author{Kui Liu}
\email{brucekuiliu@gmail.com}
\affiliation{%
  \institution{Huawei Software Engineering Application Technology Lab}
  \city{Hangzhou}
  \country{China}
}

\begin{abstract}

Detecting vulnerability fix commits in open-source software is crucial for maintaining software security. To help OSS identify vulnerability fix commits, several automated approaches are developed. However, existing approaches like VulFixMiner and CoLeFunDa, focus solely on code changes, neglecting essential context from development artifacts. Tools like Vulcurator, which integrates issue reports, fail to leverage semantic associations between different development artifacts (e.g., pull requests and history vulnerability fixes). Moreover, they miss vulnerability fixes in tangled commits and lack explanations, limiting practical use. Hence to address those limitations, we propose \ourTool, a novel framework that leverages Large Language Models (LLMs) enhanced with Chain-of-Thought reasoning and In-Context Learning to improve the accuracy of vulnerability fix detection. \ourTool comprises three components: (1) Code Change Intention, which analyzes commit summaries, purposes, and implications using Chain-of-Thought reasoning; (2) Development Artifact, which incorporates context from related issue reports and pull requests; (3) Historical Vulnerability, which retrieves similar past vulnerability fixes to enrich context. More importantly, on top of the prediction, \ourTool also provides a detailed analysis and explanation to help security experts understand the rationale behind the decision.
We evaluated \ourTool against state-of-the-art techniques, including Pre-trained Language Model-based approaches and vanilla LLMs, using a newly collected dataset, BigVulFixes.
Experimental results demonstrate that \ourTool significantly outperforms the best-performed existing approach by 68.1\%--145.4\%.
Furthermore, We conducted a user study with security experts, showing that the analysis generated by \ourTool improves the efficiency of vulnerability fix identification.
\end{abstract}

% NEEDS fixing!



\maketitle

\section{Introduction}\label{sec:intro}
% OSS and CVD
% \sw{I edited introduciton}
 %For instance, Equifax suffered from a data breach compromising over 143 million consumers' personal information due to a missed security update, resulting in over \$650 million in losses~\cite{equifax.breach}\mike{Is there another highprofile vuln? WE used equifax for vfm, coldefunda, also vulcurator}.

Software development heavily relies on the use of open-source software (OSS).
However, failing to detect and mitigate vulnerabilities in OSS can lead to catastrophic consequences~\cite{equifax.breach}.
OSS organizations generally follow the Coordinated Vulnerability Disclosure (CVD)~\cite{oss.vulnerability.guide} model to disclose vulnerabilities.
In this model, details of vulnerabilities are publicly shared once the developers feel they have had sufficient time for remediation of the security risk.
This often causes a delay between when a commit that fixes a vulnerability (\ie \textit{vulnerability fix commit}) is integrated into the codebase and when the vulnerability and or its fix are publicly announced.
The time between fix and disclosure can provide malicious users a window of opportunity to find details of vulnerabilities and exploit them in software dependent on the OSS.
Although CVD recommends applying fixes silently to avoid leaking sensitive information about the vulnerability, the risk of exploitation remains.
Hence, OSS users have a crucial incentive to monitor integrated OSS and discover vulnerability fixes to begin the remediation process as fast as possible (\ie \textit{vulnerability fix detection}).

% \jy{@Xu, consider to reuse the challenge 1,2,3 in this para}\xy{1 and 2 is ok, but no where to show 3}
Vulnerability fix detection is a complex task and poses a significant challenge for general OSS organizations without an automated approach. Furthermore,  manually monitoring all commits across integrated OSS is highly time-consuming and costly.
To solve this, automated approaches to identify vulnerability fixes have been proposed in prior studies, which typically train a deep learning model using commit-level information. 
For instance, \emph{VulFixMiner}~\cite{zhou2021finding} trains a pre-trained language model (PLM) using code change information from commits.
In their next work~\cite{zhou2023colefunda}, they proposed an approach to detect vulnerability fixes at a finer granularity by using function-level code change information. However, those approaches only leverage code change information and neglect additional software development artifacts related to commits, making it challenging to identify the intent of code changes without sufficient context, typically for nuanced code changes (\eg adding condition check). Vulnerability fix is a complex task that is often associated with issue reports~\cite{pan2022automated}, but such information is not adequately utilized in existing methods~\cite{zhou2021finding,zhou2023colefunda}. The only work that leverages development artifacts (\ie issue report) is  Vulcurator~\cite{nguyen2023vffinder}, however, it used a voting mechanism without leveraging the semantic association between artifacts. Previous approaches also fail to identify vulnerability fixes within a tangled commit --- those with a mixture of code modifications for various purposes (\eg refactoring)~\cite{barnett2015helping}. 
The limitations of existing approaches mentioned above cause them to miss many true vulnerability fix commits (evidenced by very low recalls (0.06 to 0.26) as shown in \Cref{sec:rq1_result}). 
More importantly, identifying vulnerability fixes requires specialized security expertise and project-specific domain knowledge~\cite{zhou2019devign,liu2020cd}, prior approaches only provide a prediction, without an explanation behind this decision, which hinders the use of such approaches in practice.

Recently, Large Language Models (LLMs) have demonstrated promising results in code-related tasks, such as code understanding~\cite{achiam2023gpt, leinonen2023comparing} and vulnerability understanding~\cite{fang2024teams, islam2024llm}. Intuitively, vulnerability fix detection is a task that requires the abilities of natural language understanding and code comprehension, typically in vulnerability understanding. Hence, the strong natural language and code understanding capabilities of LLMs fit the requirements of this task well.

%\sw{where we use COI?}\mike{do you mean chain of thought? It should be in the Code change intention component (1st). In Context Learning can be attributed to 2nd, 3rd, and 4th components. I think 4th is ok, thoughts @xuyang @wenhan / change if disagree}\wz{we use CoT as the prompt design style, and its been used throughout most if not all LLM queries.}
Therefore, to tackle the problem and overcome existing limitations (\ie neglecting information in development artifacts, tangled commits, and lack of sufficient explanation) of previous works, we propose \textit{\ourTool}, a novel framework that enhances vulnerability fix detection by leveraging multiple sources of information distilled by LLMs, consisting of four components.
The Code Change Intention (CCI) component analyzes a commit to extract its summary, purpose, and implications, using Chain-of-Thought (CoT) reasoning.
The Development Artifact (DA) component utilizes information from related development artifacts, such as issue reports (IRs) and pull requests (PRs), to provide additional context and enhance understanding of the commit.
The Historical Vulnerability (HV) component retrieves similar vulnerability fix commits from historical vulnerability data to further enhance the context of the commit.
Last, the Comprehensive Analysis and Vulnerability Fix Detection (CAVFD)
% \mike{previously the name of our tool was here.
% If we decide to abbreviate this component include it here} 
component combines the information distilled from the previous components to enable In-Context Learning.
The synthesized information is used to generate a final prediction, while also providing a detailed explanation in analysis to help security experts understand the rationale behind the decision.

%To the best of our knowledge, we are the first to propose a large language model (LLM) based approach on vulnerability fix detection.
%Compared to PLMs used in existing work, LLMs offer a larger context input allowing them to receive more information, and generally show superior capabilities in software engineering related tasks~\cite{hou2023large, fan2024exploring, zhou2024out,islam2024llm,kulsum2024case,cheng2024llm}.

% Our approach is based on the combination of Chain-of-Thought (COT)~\cite{wei2022chain} and In-Context Learning.
% We designed three components to provide context information to the LLM as an aid for vulnerability fix detection.
% The first component --- Code Change Intention (CCI) --- is designed to summarize the code change in \emph{3 aspects} capturing the intentions of the code change in natural language.
% The second component, Development Artifact (DA), summarises issue reports (IR) and pull requests (PR).
% And the last component, Historical Vulnerability (HV), is a RAG system designed to retrieve similar vulnerability fixes from historical vulnerability data.
% We craft the final prompt used to query the LLM with the context information from the three components in CoT style.

We evaluated \ourTool on a newly created multi-language vulnerability fix dataset which we call the BigVulFix dataset. This dataset consists of 1,689 vulnerability fix commits after 2023 collected from the National Vulnerability Database (NVD)~\cite{nvd.homepage}.
% \sw{how about the data before 2023?} \xy{VF data before 2023 are used for RAG}
\ourTool is evaluated on 6 LLMs spanning 3 LLM families in varying parameter sizes (\ie 7B--236B). \ourTool outperforms PLM-based approach consistently in terms of MCC, F1-score, and recall. For instance, \ourTool outperforms the best-performed PLM-based approach VulCurator by 68.1\%--145.4\% in terms of F1-score when using different LLMs as the base model. Our framework demonstrates performance gains ranging from 12.7\%--105.6\% over its vanilla variant, with smaller models generally benefiting more compared to their larger counterparts.
In addition, the conducted ablation analysis shows that all three components are crucial in contributing to the final performance of \ourTool. Through a user study involving security experts, we find in 80.0\% of the cases \ourTool's analysis helps security experts to understand the code change and improve the efficiency in identifying whether the change is a vulnerability fix commit.
We also conducted a failure case analysis to understand the limitations of \ourTool, and provide insights for future work.

In summary, this paper makes the following contributions:
\vspace{-0.06in}
\begin{itemize}
    \item To the best of our knowledge, we are the first study to investigate vulnerability fix detection using LLM.
    \item We propose a novel framework, \ourTool, that enhances vulnerability fix detection by leveraging multiple sources of information based on LLMs. 
    \item We conducted extensive evaluation on \ourTool, including an ablation study to evaluate the contribution of each component in \ourTool and a user study to evaluate the usefulness of \ourTool's analysis result for security experts.
    \item We collected and release a new dataset, BigVulFixes, containing 1,689 vulnerability fix commits from 7 major programming languages after 2023 to avoid data leakage with LLM's pre-training data.
    \item We perform a bad case analysis on the limitations of our approach to promote future work on vulnerability fix detection.
\end{itemize}

\section{Background and Related work}
In this section we introduce the background of vulnerability fix detection and LLMs.

\subsection{Vulnerability Fix Detection}
A typical vulnerability fix detection task takes a commit as input and outputs whether the commit is fixing a vulnerability.
Existing approaches focusing on this problem mostly leverage PLM techniques through some kind of embedding technique to capture code change information.
% first work on the topic
The first work on vulnerability fix detection is introduced by~\citet{zhou2017automated}.
They used commit messages and bug reports to automatically identify vulnerability fix commits.
% second work
\citet{chen2020machine} later explored the topic with the idea of vulnerability-relatedness, to capture how each commit can be related to addressing a vulnerability.

During this time, trying to associate a commit with a vulnerability was mainly to curate vulnerability-related information to aid data collection and cleaning.
Having high-quality data for vulnerability fixing code is critical for other vulnerability tasks such as automated vulnerability repair.
\citet{zhou2021finding} further explored the idea of detecting vulnerability fix commit especially in the case of silent fixes where the vulnerability fix information may not be publicly disclosed.
\citet{sabetta2018practical} leveraged an SVM based approach with features from commit messages and patch information to predict whether a commit is security related.
\citet{nguyen2022hermes} combined three classifiers each for the patch, commit, and associated issue content to create a joint classifier to identify vulnerability fix commits.
\citet{xu2017spain} proposed SPAIN to identify vulnerability fixes at the binary level.
Their work is especially useful when the target software is not available in source code and is distributed only in binary.

% granularity
There have also been efforts in the direction of a more fine-grained level of vulnerability detection.
For example, \citet{zhou2023colefunda} proposed a framework to analyze vulnerable code at the function level using deep learning techniques.
Their framework can identify vulnerability fixes, CWE category, and exploitability.
% graph-based
Alternative representations of the changed code have also been explored, especially using graph-based techniques.
\citet{nguyen2023vffinder} proposed a graph-based approach for detecting silent vulnerability fixes.
They construct graphs using the AST before and after a commit.
Their work show significant improvement in performance in real-world C/C++ projects.

\subsection{Large Language Models}
Popularized by Generative Pretrained Transformer (GPT) models~\cite{ouyang2022training}, LLMs have shown great potential in software engineering tasks.
Recent work explores the capabilities of LLMs in solving several unique software engineering problems ~\cite{hou2023large, fan2024exploring, zhou2024out,islam2024llm,kulsum2024case,cheng2024llm}.
In practice, LLMs are generally tailored for use in domain-specific tasks through prompt engineering and/or fine-tuning.

Prompt engineering is an essential step in interacting with an LLM to alter its response.
The characteristics of the prompt (\eg vocabulary, style, tone) can greatly affect the generated response of the LLM~\cite{zamfirescu2023johnny}.
Carefully crafted prompts can help improve the capability of LLMs in specific tasks.
For example, CoT~\cite{wei2022chain} is a common form of prompt engineering that separates the prompt into smaller, individual steps, which improves the reasoning capabilities of LLMs.
LLMs also have varying context lengths which limit the length of the prompt, ultimately truncating information that exceeds this limit.
Thus, prompt engineering also generally aims to tailor prompts to become brief by condensing information.
In addition, prompt engineering is an effective way to work around the knowledge cutoff of LLMs.
LLMs are typically trained with information that precedes a certain knowledge cutoff date.
When LLMs are queried for information beyond the cutoff date or not in the training dataset, they often hallucinate and create unsuitable answers.
An example of this often arises from 0-shot prompting strategies~\cite{kojima2022large}, which prompt LLMs to perform tasks that they are not explicitly trained in.
This issue is often resolved by providing contextual information in the prompt.
A canonical solution is In-Context Learning (ICL)~\cite{dong2022survey}, which includes examples and/or demonstrations of the task in the prompt. 
% Such approaches often include Retrieval Augmented Generation (RAG)~\cite{lewis2020retrieval}, which is used to provide the LLM with context information during querying.

With its superior capability in code related tasks such as code understanding~\cite{shen2022benchmarking}, code summary~\cite{wang2023codet5+}, and code generation~\cite{bareiss2022code,gilbert2023semantic,jiang2023selfevolve,liu2024your}), LLM overcomes many limitations of previous techniques~\cite{wu2023effective,hou2023large}.
The strength of LLMs leads to researchers adopting the tool for vulnerability related tasks.
Most of the previous work has focused on vulnerability detection~\cite{zhou2024large}.
\citet{chan2023transformer} proposed a system based on vulnerable code patterns to detect vulnerabilities in code.
\citet{thapa2022transformer} researched LLM's performance of C/C++ source codes with multiple vulnerabilities.
\citet{cheng2024llm} presented an approach Vercation to identify vulnerable versions of open-source C/C++ software.
\citet{du2024vul} proposed Vul-RAG, a framework that first constructs a vulnerability knowledge base that contains CVE information. The framework is then able to predict whether a given code snippet is vulnerable by an RAG system that retrieves from the knowledge base.

% To our best knowledge, we are the first work in leveraging LLM on the vulnerability fix detection task.

\section{Challenges and Motivation}\label{sec:callenges}
In this section, we outline the primary challenges faced by current vulnerability fix detection techniques and discuss the types of context information that could be leveraged to address them.

\subsection{Challenge \#1: Tangled Commits}

\begin{figure}[]
    \centering
    \includegraphics[width=0.6\columnwidth]{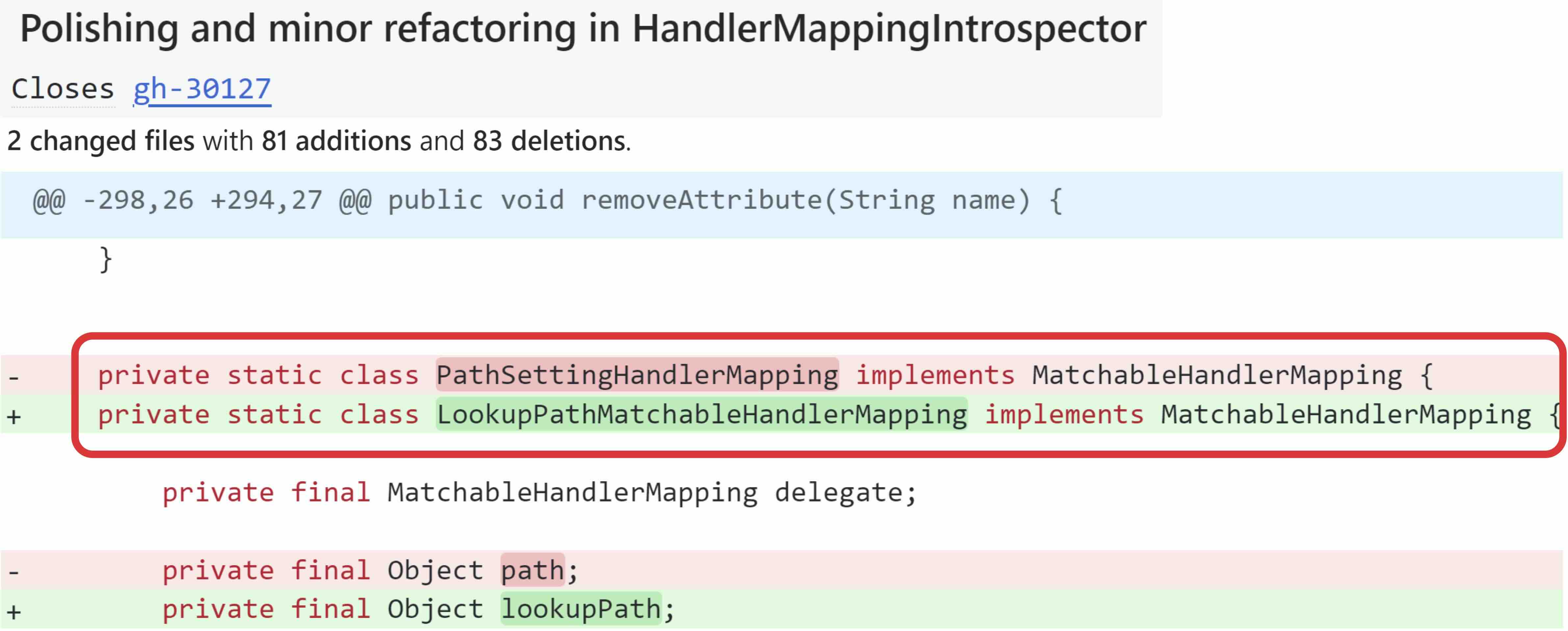}
    \caption{An example of tangled commit with 164 lines changed, while only two lines (in red box) are related to vulnerability fix.~\cite{github.spring.202fa5c}}
    \label{fig:motivation_1}
\end{figure}

One of the key challenges in detecting vulnerability fixes is identifying security-related changes within a commit that contains multiple modifications for various purposes, such as feature improvements, refactoring, and vulnerability resolution~\cite{barnett2015helping}. This mix of intentions complicates the task for traditional methods~\cite{herzig2013impact}, particularly those that rely solely on analyzing code changes to detect vulnerabilities~\cite{zhou2021finding,zhou2023colefunda}. 
For example, in \Cref{fig:motivation_1}, only two lines out of 164 in the commit address a vulnerability by altering the initialization and usage of \texttt{LookupPathMatchableHandlerMapping}, while the rest relates to refactoring and feature adjustments. The substantial amount of non-vulnerability-related changes in such tangled commits introduces noise, making it difficult for existing approaches to accurately detect the real vulnerability-related fixes. In fact, none of the current methods~\cite{zhou2021finding,zhou2023colefunda,nguyen2023vffinder} can identify this particular case.

To tackle this issue, we interviewed security experts to understand how they discern vulnerability fixes, especially in tangled commits. From their insights, we identified three crucial aspects of information needed to determine a vulnerability fix: the summary of the code change, the purpose behind the changes, and their potential implications. By concentrating on these aspects rather than just the raw code changes, we can effectively filter out noise from unrelated modifications, thereby enhancing the accuracy of models for vulnerability fix detection.
% focusing on these aspects helps filter out noise from unrelated modifications and improve the accuracy of vulnerability-fix detection.

% \label{motivating-example}

%\sw{we only present the challenges, however, did not motivate which part of information could be leveraged to address this? we need to explicitly point out here. }
%\xy{added a brief 2-sentence description of how we use CCI to address this challenge.}\sw{do not introduce the component of our approach. we should introduce what information would be used }
%In our paper, we design the Code Change Intention (CCI) component\wz{need to expand since this is the first place we introduced the term} to leverage the information within the commits together with the reasoning ability of LLM. To achieve that, we combine the commit data with a guided prompt template that uses a Chain-of-Thought (CoT) approach to systematically extract and distill the security-relevant information. We describe the details of the CCI component in~\Cref{sec:CCI}.

\subsection{Challenge \#2: Insufficient Information Within Commits}

\begin{figure}[]
    \centering
    \includegraphics[width=0.8\columnwidth]{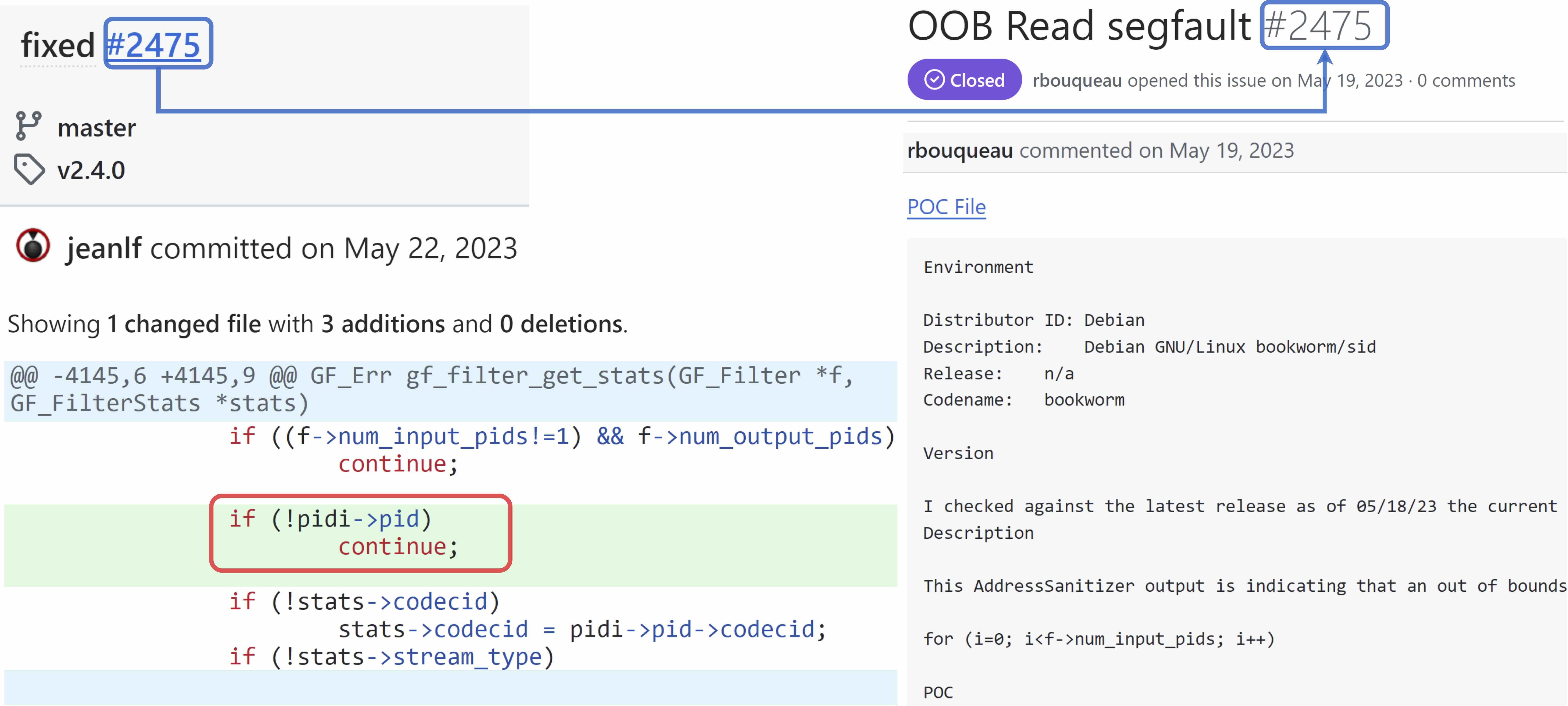}
    \caption{An example of a commit~\cite{github.gpac.c88df2e} with only 2 lines changed (in red box), while related issues reports~\cite{github.gpac.issues.2475} (in blue box) provided critical information.}
    \label{fig:motivation_2}
\end{figure}

In many cases, commit messages and code changes alone do not provide enough information to determine whether a commit is related to a vulnerability fix.
This challenge is particularly pronounced when changes are subtle and or the commit message is vague, making it difficult for traditional methods~\cite{zhou2021finding, zhou2023colefunda} to accurately identify vulnerability fix commits.
For example, the commit shown in~\Cref{fig:motivation_2} simply adds three lines of code to \texttt{filter\_session.c} with a conditional check.
The commit message, ``fixed \#2475,'' provides little insight into the nature of the changes or whether they are security relevant.
Hence, based on the code and commit message alone it is nearly impossible to identify whether this is a vulnerability fix or a routine update, even for human experts.
However, when we investigate the related development artifact --- issue report \#2475 --- it becomes evident that this commit addresses a security vulnerability.
The issue report details a vulnerability involving improper handling of certain filter parameters, which could lead to a security flaw causing an out-of-bounds read and segmentation fault.
The commit directly resolves this issue, though without the context provided by the issue report, commit-only methods are likely to miss this crucial connection and result in a false negative prediction.
Therefore, we aim to leverage related development artifacts to enrich commits for vulnerability fix detection.

\label{motivating-example}

%In our paper, we design the Development Artifact (DA) component to address the lack of context within commits by integrating external development artifacts such as issue reports and pull requests. We combine the artifact data with a structured prompt template similar to the ICC component to extract relevant information to provide a more comprehensive context. We described the details of the DA component in \Cref{sec:DA}.

\subsection{Challenge \#3: Missing Project-Specific Contextual Information}
%\sw{I would say, only with the code change without knowing the the project-specific context information, it is challenging to differentiate between vul fix and non-vul fix. as shown in the example, the change related to a condition check is challenging to figure out. However, project typically have historical data that have similar change patterns related to vul. for example, in the history, we found... }\xy{modified below, please check}
% 1st commit: https://github.com/protobufjs/protobuf.js/commit/e66379f

\begin{figure}[]
    \centering
    \includegraphics[width=0.9\columnwidth]{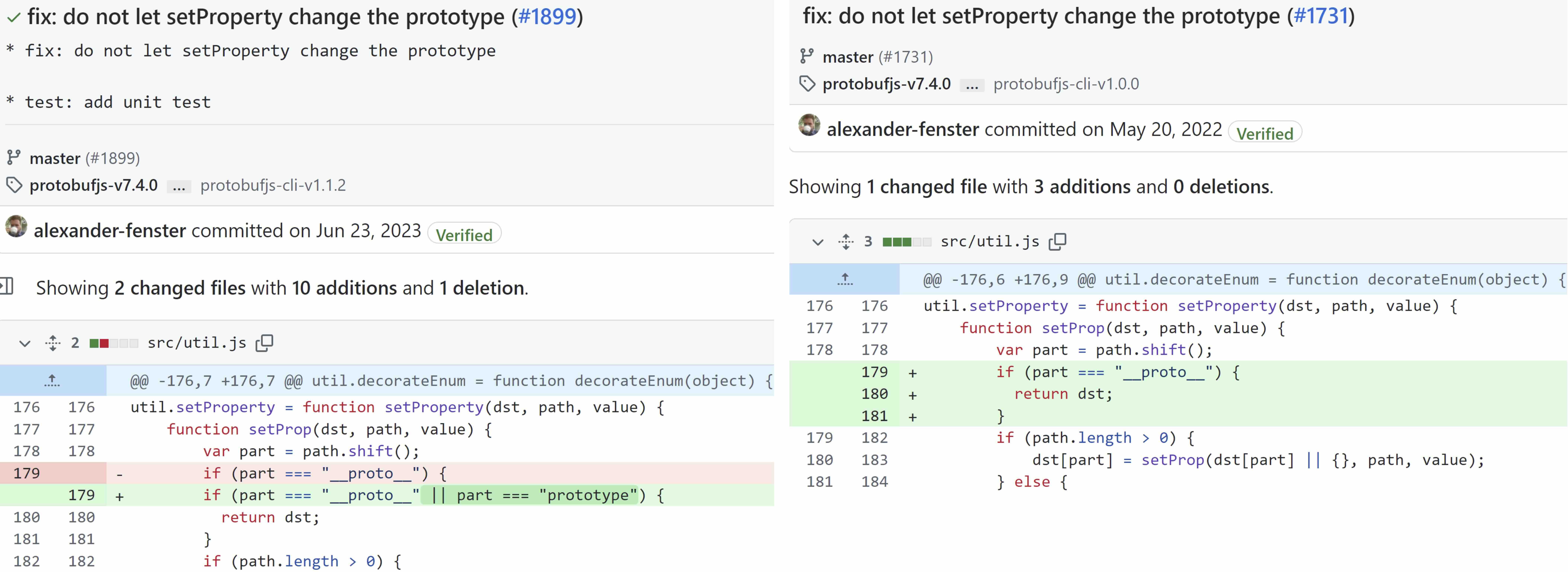}
    \caption{A commit only with small change by adding condition check (left~\cite{github.protobufjs.e66379f}) and its relevant historical vulnerability fix commit (right~\cite{github.protobufjs.3357ef7}).}
    \label{fig:motivation_3}
    \vspace{-0.2in}
\end{figure}

It is usually challenging to predict only based on the code changes without a deeper understanding of the corresponding OSS software (i.e., project-specific domain knowledge).
For instance, a commit shown on the left in~\Cref{fig:motivation_3} includes a small code change that adds checks in the \texttt{setProperty} function to prevent changes to the prototype property --- a common source of prototype pollution vulnerabilities.
While the commit message uses the word ``fix'', the simplicity of the change makes it difficult to determine whether this is a vulnerability fix, typically for traditional methods, which rely on a surface-level analysis of the commit message and or code~\cite{zhou2021finding, zhou2023colefunda, nguyen2023vffinder}, may struggle to differentiate between them without such information.
Conveniently, projects typically have historical commits that possibly contain similar fixes can provide more context-specific for the project.
For instance, in ~\Cref{fig:motivation_3} (right), we present a historical vulnerability fixing commit, which includes similar changes to the same function in a prior commit, where the addition of a check for the \texttt{\_\_proto\_\_} property was introduced to mitigate a known prototype pollution vulnerability. 
This example shows that historical vulnerability fixes can be leveraged to supply the missing context needed to identify new vulnerability fix commits.

%In our paper, we design the Historical Vulnerability (HV) component with a RAG to addresses the challenge of ambiguity in commit data by retrieving similar past vulnerability-fix commits from a vector database. By comparing the new commit with historical vulnerabilities, the HV component provided extra reference to help identify patterns and security-relevant changes.

\section{Methodology}\label{sec:method}
\begin{figure}[]
    \centering
    \includegraphics[bb=0 0 2236 1133, width=1\linewidth]{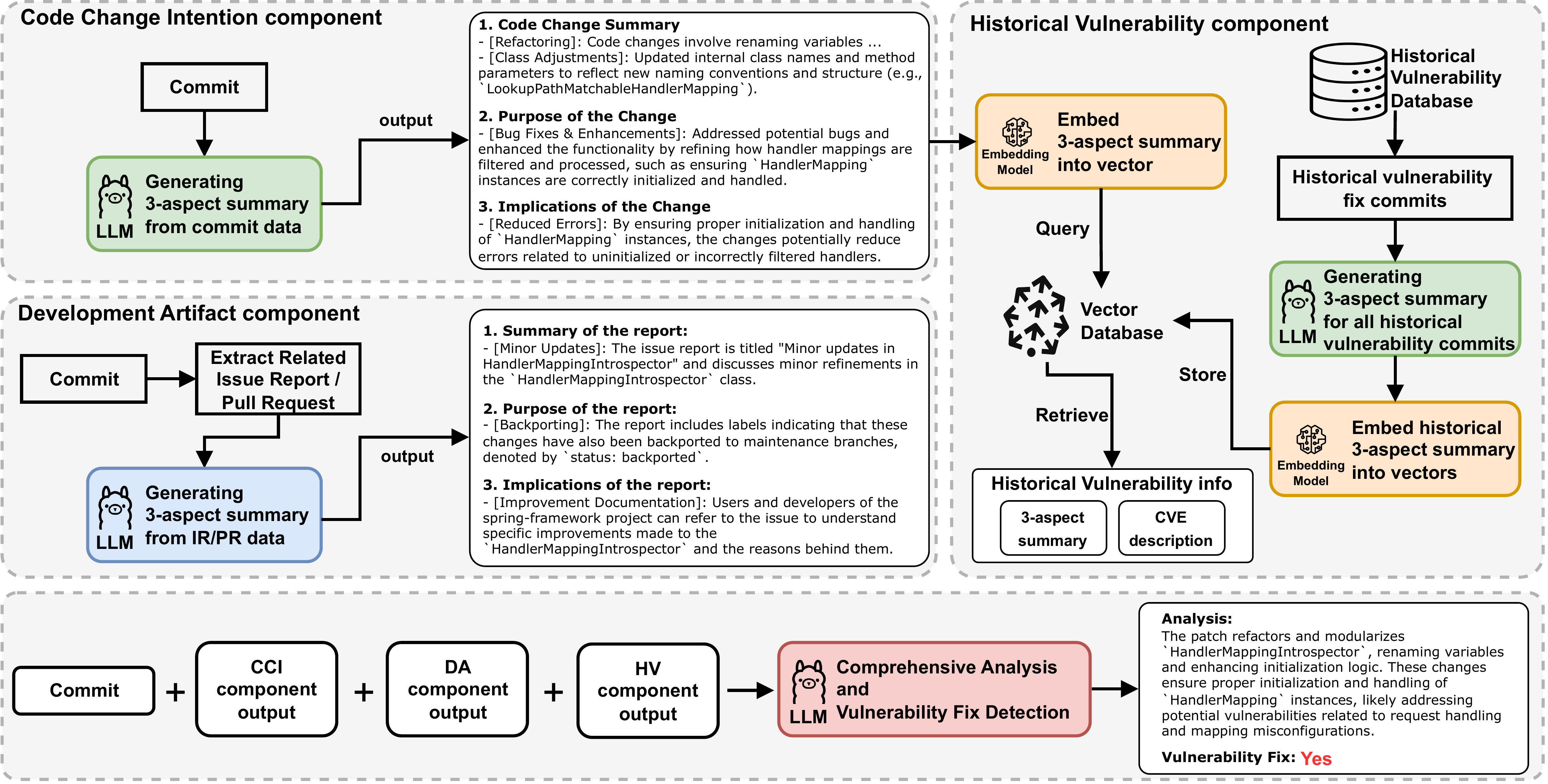}
    \caption{The framework of \ourTool.}
    \label{fig:framework}
    \vspace{-0.2in}
\end{figure}

To address the challenges discussed in Section~\ref{sec:callenges}, we present \ourTool, which leverages information that is distilled from multiple sources by leveraging LLMs to enhance the vulnerability fix detection. \Cref{fig:framework} provides an overview of our framework.
\ourTool consists of four components: Code Change Intention (CCI), Development Artifact (DA), Historical Vulnerability (HV), and Comprehensive Analysis and Vulnerability Fix Detection (CAVFD). 

More specifically, to address Challenge \#1, the CCI component leverages LLMs with Chain-of-Thought (CoT) techniques to construct a summary of the commit that focuses on three crucial aspects (\ie code change summary, purpose of the change, and implications of the change). To address Challenge \#2, the DA component leverages the information extracted from related development artifacts (\ie issue reports (IR) or pull requests (PR) in our case) to provide additional information to enrich the context for a commit. To address Challenge \#3, the HV component builds a vector database and retrieves similar vulnerability fix commits from historical vulnerability data. The information from these components is then combined to enhance the analysis in the CAVFD component, where a comprehensive prompt template is employed to guide the LLM through all the relevant aspects of the commit. This structured prompt not only predicts whether the given commit is a vulnerability fix, but more importantly, it generates an in-depth explanation in analysis to assist security experts in understanding the rationale behind the prediction.

%\sw{highlight that on top of prediction, we also provide the analysis to help security experts in identifying vulnerability fix}.\xy{fixed, please check}

As highlighted in \Cref{fig:framework}, we provide a running example to illustrate the procedure of \ourTool. The outputs are taken from the commit from Challenge \#1, which by leveraging our approach, can be correctly identified as a VF commit.

%\sw{use one sentence to summarize each component, what to do, with what technique.change IRPR to IR and PR} \sw{I don't think we can say it is rag, since rag is related to generation, i think our only do retrieval?}\xy{this is like a regular rag when we consider the generation part is in the final prompt}\sw{so HV itself is not RAG, right?}

\subsection{Code Change Intention (CCI)}\label{sec:CCI}

%\sw{I recall you fine-tuned a model to extract information?}\xy{we used to have a codellama model for that, now we use prompt only}

The Code Change Intention component takes raw commit data (\ie the code diff and commit message) as input, and outputs a structured summary that abstracts the intention behind the commit.
However, obtaining this information is not trivial, as the commit data does not explicitly provide this level of structured reasoning.
To generate such information, we leverage and enhance the reasoning ability of LLMs using CoT techniques. Specifically, we design a structured prompt that breaks down the reasoning process into steps, mimicking how a security expert would analyze the commit.
%\sw{what is the reason of have key point and optional point,m make it clear. does it mean only at most two points will be listed by LLM}\xy{added description below}
The prompt template, as shown in~\Cref{box:code_change_intention_prompt}, guides the LLM to analyze the commit code changes and distill relevant information into three key aspects (referred to as \textit{3-aspect summary}): the summary, the purpose, and the implications as below. 
% \sw{I changed the name, apply them throughout the paper.}\xy{get, fix when see}
\begin{enumerate}
    \item \textbf{Code Change Summary.} In the first step of the prompt, we instruct the LLM to focus on identifying the primary action of the code change. The goal here is to abstract the core modification and summarize different kinds of code changes.
    \item \textbf{Purpose of the Change.} The second step of the prompt instructs the LLM to reason through why the change was made. This step categorizes the commit into broader categories such as refactoring, feature enhancement, or fixing a vulnerability.
    \item \textbf{Implications of the Change.} In the final step, the prompt asks the LLM to consider the broader consequences of the change, including its potential impact. This step ensures that any security-related modifications are properly captured and that the potential risks or fixes introduced by the commit are fully understood.
\end{enumerate}

In addition, to ease the extraction process for further components, we provide concrete instructions with a demonstration example to ensure LLM outputs the 3-aspect summary in the format we expect. Note that one commit, such as tangled commits, probably has multiple points of summary, purposes, and implications. We instruct LLM to generate multiple points (i.e., Key Point/Optional Key Points) for each aspect if it is applicable.
%\xy{added more detail to above prompt design description}
% \begin{tcolorbox}[colback=yellow!10!white, colframe=white, sharp corners]
% \small\textbf{System Prompt:} You are a helpful software developer assistant specializing in software development life-cycle to help other developers understand the characteristics of software patches.
% \\
% \small\textbf{User Prompt:} You are given the following software patch: \textit{\textbf{\{Commit\}}}
% \\Think step by step and provide an analysis describing the following characteristics.
% \\1. Code Change Summary\\2. Purpose of the Change\\3. Implications of the Change
% \\Provide the analysis in bullet point format for each characteristic. Each bullet point should start with a key point and then briefly describe a main idea or fact from the text. Ensure each point is concise and captures the essence of the main idea it's summarizing. Here is an example of the desired format:
% \\1. Code Change Summary
% \\- [Key Point]: <description>
% \\- [Optional Key Point]: <description>
% \\2. Purpose of the Change
% \\- [Key Point]: <description>
% \\- [Optional Key Point]: <description>
% \\3. Implications of the Change
% \\- [Key Point]: <description>
% \\- [Optional Key Point]: <description>
% \end{tcolorbox}

\prompt{Prompt Template: Code Change Intention}{
\textbf{System Prompt:} You are a helpful software developer assistant specializing in software development life-cycle to help other developers understand the characteristics of software patches.
\\\textbf{User Prompt:} You are given the following software patch: \textit{\textbf{\{Commit\}}}
\\Think step by step and provide an analysis describing the following characteristics.
\\1. Code Change Summary\\2. Purpose of the Change\\3. Implications of the Change
\\Provide the analysis in bullet point format for each characteristic. Each bullet point should start with a key point and then briefly describe a main idea or fact from the text. Ensure each point is concise and captures the essence of the main idea it's summarizing. Here is an example of the desired format:
\\1. Code Change Summary
\\- [Key Point]: <description>
\\- [Optional Key Point]: <description>
\\2. Purpose of the Change
\\- [Key Point]: <description>
\\- [Optional Key Point]: <description>
\\3. Implications of the Change
\\- [Key Point]: <description>
\\- [Optional Key Point]: <description>
}
{box:code_change_intention_prompt}{The prompt template of Code Change Intention.}

\subsection{Development Artifact (DA)}\label{sec:DA}

In the DA component, we aim to integrate the information of external development artifacts such as issue reports (IRs) and pull requests (PRs) to enrich the analysis of a given commit. However, these artifacts can be lengthy and include irrelevant details, such as issue report templates and CI/CD notifications. To address this, and inspired by the CCI component in~\Cref{sec:CCI}, we design a structured prompt that guides the LLMs through the process of analyzing and summarizing the related IRs and PRs to generate a 3-aspect summary which is similar to that for a commit. More specifically, the DA component generates a concise summary that abstracts the intention behind these artifacts by focusing on three core aspects: the summary, the purpose of the changes, and the potential implications. The prompt template is shown in Figure~\ref{box:irpr_summary_prompt}.

\prompt{Prompt Template: IR/PR Summary}{
\textbf{System Prompt:}You are a helpful software developer assistant specializing in software development lifecycle to help other developers understand characteristics of software components such as patches, issue reports, pull requests, etc.
\\
\textbf{User Prompt:} You are given the following Github issue report title and body information in JSON format which is related to a commit:\textit{\textbf{\{Commit\}}}
\\Think step by step and provide an analysis describing the following characteristics.
\\1. Summary of the report\\2. Purpose of the report\\3. Implications of the report
\\Provide the analysis in bullet point format for each characteristic. Each bullet point should start with a key point and then briefly describe a main idea or fact from the text. Ensure each point is concise and captures the essence of the main idea it's summarizing. Include 1-3 key points. Here is an example of the desired format:
\\1. Summary of the report:
\\- [Key Point]: <description>
\\- [Optional Key Point]: <description>
\\2. Purpose of the report:
\\- [Key Point]: <description>
\\- [Optional Key Point]: <description>
\\3. Implications of the report:
\\- [Key Point]: <description>
\\- [Optional Key Point]: <description>
}
{box:irpr_summary_prompt}{The prompt template of Development Artifact.}

% This output, termed "Development Artifact," encapsulates the essential context needed to understand the role of the commit within the broader development process. By summarizing these artifacts, we ensure that the LLM has access to a comprehensive understanding of the commit’s purpose and implications, which is often crucial for accurately identifying vulnerability fixes. The inclusion of development artifacts provides an additional layer of context, enabling a more holistic analysis that goes beyond the code itself.
% By abstracting the information, we avoid the pitfalls of traditional code search techniques, which might be skewed by unchanged code segments, and instead focus on the intent and impact of the change. This abstraction is critical because existing studies on searching code often fail to capture the full semantics of code changes. The unchanged portions of the code might influence the search results in misleading ways, as they do not reflect the purpose of the change. Our approach mitigates this issue by focusing on the summarized intent and implications, which are more relevant for identifying vulnerabilities.

\subsection{Historical Vulnerability (HV)}
%\sw{name it as component, do not name it as data.}}\sw{too wordy on motivation, focus on methodology}\xy{moved some part to motivation example}\mike{should the names be plural? component vs components? or just exclude "component" from the subsection title}

%\mike{I edited this paragraph, old one is commented below if needed}
In this component, we aim to leverage information from historical vulnerability fix commits. More specifically, given a commit, HV aims to retrieve similar vulnerability fix commits to enrich the given commit.
As discussed in \Cref{sec:callenges}, the vulnerability fix commit can be multi-purpose and contain code addressing other issues.
Therefore, we decide to retrieve similar vulnerability fix commits based on their intention. To do this, we need to construct a database of historical vulnerability fixes. 
First, we collect historical vulnerability fix commits from existing vulnerability databases such as NVD~\cite{nvd.homepage}. We generate the 3-aspect summary for all the collected vulnerability fix commits using the CCI component.
We then embed and vectorize the generated three-aspect summary for each vulnerability fix commit using a sentence embedding model~\cite{li2023towards}, and store them in a vector database.
Alongside the generated summaries, we also store their corresponding vulnerability metadata, including the CVE ID and CVE description.

When processing a new commit, we begin by using the CCI component to obtain a 3-aspect summary for the commit. We search for the nearest instance from historical vulnerability database, based on the similarity of their 3-aspect summary. With the nearest instance, we gather the 3-aspect summary and CVE description from its metadata as the output of the HV component.

\subsection{Comprehensive Analysis and Vulnerability Fix Detection (CAVFD)}

%\sw{need more justification why the prompt is designed so. note that the prompt is our ALGORITHM, so we need more explanation and focus on this, instead of repeating motivation in methodology}

In the final step, we combine multiple sources of output collected from the three components (\ie CCI, DA, and HV), together with the commit’s code change and commit message, into a comprehensive prompt for the LLM. This step ensures that the model has access to all relevant characteristics of the commit, not only in terms of the code itself, but also the surrounding context and historical vulnerability fix commits. The output is a prediction with a structured analysis and reasoning on how the decision is made.

The prompt template is shown in \Cref{box:vf_prompt}. The prompt is designed to follow a multi-dimensional approach~\cite{chen2023unleashing} to ensure a thorough, structured analysis for a given commit.
We do this by guiding the LLM to analyze and integrate information from the components using the CoT and ICL techniques before making its decision.
We first provide the raw commit data (code diff and commit message) as the patch content, following by the output from the three components. 
Next, we design the prompt to include a two-step task for the LLM: Comparison and Analysis.
In the Comparison phase, we prompt the LLM to evaluate the current patch against the retrieved historical fixes, to avoid potential bias and ensure an evidence-based analysis. This is because we cannot ensure that the retrieved historical vulnerabilities are actually relevant to the current commit.
In the Analysis step, we ask the LLM to synthesize information from the three components to determine whether the patch is a vulnerability fix, and to provide justification.
Finally, we instruct the LLM to generate the response in JSON format including its detailed analysis and finally its decision. The analysis process is designed to output an analysis that can assist the user of \ourTool in the manual screening process to help them better understand the decision-making process of the LLM. We also conduct a user study in~\Cref{sec:rq3} to evaluate the usefulness of the resulting generated analyses.
% \begin{tcolorbox}[colback=yellow!20!white, colframe=white, sharp corners]
% This is a code snippet with a vulnerability \texttt{[CVE ID]}: \texttt{[Vulnerable Code]}. The vulnerability is described as follows: \texttt{[CVE Description]}. The correct way to fix it is by \texttt{[Patch Diff]}. The code after modification is as follows: \texttt{[Patched Code]}. Why is the above modification necessary?
% \end{tcolorbox}
% This holistic approach, which leverages the Chain-of-Thought and In-Context Learning techniques, ensures that the LLM not only understands the immediate code change but also situates it within a broader context of historical vulnerabilities and related development activities. By synthesizing information from multiple sources, our methodology significantly enhances the accuracy of vulnerability-fix detection, making it a valuable tool for improving software security in open-source projects.

\prompt{Prompt Template: Comprehensive Analysis and Vulnerability Fix Detection}{
\textbf{System Prompt:}You are a helpful software developer assistant specializing in vulnerability detection to help other developers understand characteristics of software patches and discover potential vulnerabilities.
\\
\textbf{User Prompt:} You are given the following details for analysis:
\\1. Patch Content: \textit{\textbf{\{Commit\}}}
\\2. Related Issue Report / Pull Request Summary: \textit{\textbf{\{DA component output\}}}
\\3. Three Aspect Analysis of the Patch: \textit{\textbf{\{CCI component output\}}}
\\4. Similar Historical Vulnerability Fix Information: \textit{\textbf{\{HV component output - CVE description\}}}
\\5. Three Aspect Analysis of the Historical Vulnerability Fix: \textit{\textbf{\{HV component output - 3-aspect summary\}}}
\\ Task:
\\1. Comparison:
\\- Carefully compare the current patch with the historical vulnerability fix to avoid bias.
\\- Ensure that you consider the similarities and differences highlighted in the three aspect analyses.
\\2. Analysis:
\\- Use the information from the Related Issue Report / Pull Request Summary to understand the context and motivation behind the patch.
\\- Determine whether the current patch is intended to fix a vulnerability. You must provide evidence if you think its a vulnerability fix.
\\Your output should follow below syntax:
% \\ \{
\\ \ \{"analysis": "<Detailed analysis of whether the patch is to fix a vulnerability>",
\\ \ "vulnerability\_fix": "<yes\ or\ no>"\}
% \\ \}
}
{box:vf_prompt}{The prompt template of Comprehensive Analysis and Vulnerability Fix Detection.}

\section{Experimental Settings}
In this section, we present research questions (RQs), datasets, evaluation metrics, our analysis approach for RQs, and implementation details.

\subsection{Research Questions}
We evaluate \ourTool in different aspects to answer the following research questions.

\begin{itemize}
    \item\rqone
    \hfill
    \item \rqtwo
    \hfill
    \item \rqthree
    \hfill
    \item \rqfour
    %\hfill
    %\item \rqfive

\end{itemize}
\subsection{Data Collection}
\subsubsection{Date Range Selection}
LLMs are trained on extensive data, which results in a knowledge cutoff date reflecting the most recent information they possess. For our task, feeding LLMs with historical vulnerabilities predating their knowledge cutoff can lead to data leakage, as the model might already be aware of these vulnerabilities. As a mitigation, we restrict our analysis to vulnerabilities after 2023 to ensure that the data used is post-knowledge cutoff and reduces the risk of data leakage.

\subsubsection{Vulnerability and Non-vulnerability Fix Commit Selection}
We begin the data collection process by collecting historical CVE data from NVD~\cite{nvd.homepage}.
NVD contains a vast amount of vulnerability data covering numerous open-source software (OSS) that cover a large spectrum of development activities and purposes.
We only include CVEs that possess GitHub commit URLs in their references, which indicate a vulnerability fix.
To obtain this info, we use GitHub's REST API endpoint to directly access repository and the language key in the response JSON~\cite{github.api}.
To avoid the long tail effect from the diversity of vulnerabilities, we limit ourselves to vulnerabilities from 7 programming languages, namely Java, C, C++, Rust, JavaScript, Python, and Go.

In real-world open-source software (OSS) development, VF commits are exceptionally rare, comprising only a tiny fraction of total commits. The ratio of VF to normal commits can be as low as 1 in 1,000. For example, in the OSS project FFmpeg, we collected 114,210 total commits, of which only 124 were VF commits (0.1\%). This extreme class imbalance makes our task significantly more challenging than typical binary classification tasks, which often assume a more balanced distribution (close to a 1:1 ratio). 
To address this imbalance and reduce the number of NVF commits evaluated, we followed the sampling strategy from the Big-Vul dataset~\cite{fan2020ac} and selected a ratio of 1:16 between VF and NVF commits. We randomly sample NVF commits from the same OSS where we collected vulnerability fix commits.

Our final evaluation dataset BigVulFixes consists of 1,689 VF and 26,468 NVF commits, reflecting this sampled ratio. Additionally, to handle extreme outliers and ensure compatibility with the maximum token length of the language models we use, we excluded patches longer than the 99th percentile of patch token lengths (approximately 30,000 tokens).

\subsection{IR/PR Data Collection}
For all commits gathered from the previous steps, we collect related IR/PR URLs and their information using two approaches. The first approach parses the commit message to find any references to IR/PRs. GitHub uses an autolink feature via specific formatting syntax, allowing developers to reference IR/PR information directly in commit messages~\cite{github.api}.
We design a regular expression to find such references to IR/PRs based on the defined autolink formatting syntax. The second approach accesses the ``List pull requests associated with a commit'' GitHub REST API endpoint.
This endpoint provides the pull request the corresponding commit is included in. 
This method is particularly useful for cases where developers do not reference the related pull request in the commit message, causing the first approach to fail. 
Finally, we use the GitHub REST API to retrieve the information from each of the collected IR/PR URLs. We collected a total of 17,791 IR/PR data for the BigVulFixes dataset.
% Mike; if needed or smeone complains We can argue that obtianing relations from commit to IRPR and viceversa by collecting all IRPRs for each included repo is infeasible due to the vast number of IRPRs across all repos, hence we attempt to find this relation from the commit itself, which is the main object of interest. not sure if we still include all VFs regardless of repo but that is another point
% Mike: This Github discussion thread discusses that the only way to retreive an issue from a commit is to use the parse the msg to find any formatting syntax: https://github.com/orgs/community/discussions/24541#discussioncomment-3244458

% \xy{TBD}
\subsection{Historical Vulnerability Data Collection}
In the previous section, we describe how we collect data after 2023 for evaluation, therefore, to avoid data leakage, the data range we select for the history vulnerability dataset will be all vulnerable data before 2023. Similar to the previous section, we start by collecting all history CVE information on NVD~\cite{nvd.homepage} before 2023 and collected 22,745 vulnerabilities from vulnerability advisory NVD. For each historical vulnerability, we collect its associate vulnerability fix commit and CVE description. %\sw{this is part of implementation, move to implementaiton}
% We successfully retrieve history vulnerability data for 96.5\% (50,498 out of 52,343) of the testing data.

\subsection{Implementation Detail} 
% \sw{move all footnote's URL to reference (use cite) if we need space.}
\subsubsection{LLMs selection}

\ourTool is a framework that can be implemented with any LLM and embedding models. In this study, we selected SOTA LLMs based on their performance on both code-related benchmarks (\eg HumanEval~\cite{chen2021evaluating}, MBPP EvalPlus~\cite{liu2024your}) and general benchmarks (\eg MMLU~\cite{hendrycks2020measuring}, IFEval~\cite{zhou2023instruction}). We focused on three LLM families: Llama~\cite{dubey2024llama}, Qwen~\cite{yang2024qwen2}, and Deepseek~\cite{zhu2024deepseek}. For the Llama family, we chose Llama3.1-70B and Llama3.1-8B. From the Qwen family, we selected Qwen2-70B and Qwen2-7B. For the Deepseek family, we included Deepseek-Coder-V2 (236B) and its smaller version, Deepseek-Coder-V2-Lite (16B). We use the "instruct" fine-tuning version of all LLMs.

% \sw{remove this if need space.}
We deployed Llama, Qwen and Deepseek-Coder-V2-Lite using vLLM~\cite{kwon2023efficient} on Ascend 910 NPUs. For Deepseek-Coder-V2, we utilized the official API provided by DeepSeek. The total computational cost of the experiment was approximately 2.5 billion tokens.
While more powerful LLMs like Llama3.1-405B or GPT-4o are available, they are either closed-source or too computationally intensive to deploy for our study. Therefore, we selected the models mentioned above. 
% Also, our selected LLMs are well-known and commonly used.

\subsubsection{Embedding Model and RAG implementation}
For the embedding model used in our Retrieval-Augmented Generation (RAG) to embed the 3-aspect summary, we chose gte-Qwen2-7B-instruct~\cite{li2023towards}. This model was the state-of-the-art sentence embedding model on the Massive Text Embedding Benchmark~\cite{muennighoff2022mteb} as of June 16, 2024. We use ChromaDB~\cite{chromadb} to implement the vector database to use as part of HV and RAG. When querying the HV database, 
We filter the HV query results using a condition to retrieve historical vulnerabilities only with the same programming language of the current commit.
We use the default Euclidean Distance function from ChromaDB to calculate and retrieve the most similar historical vulnerability. 

\subsection{Evaluation Metrics}

We use precision, recall, F1-score and Matthews correlation coefficient (MCC)~\cite{matthews1975comparison} as our evaluation metrics, which are widely used in previous studies~\cite{zhou2019devign,li2021vulnerability,chakraborty2021deep,rahman2024towards,yang2023does}. Unlike previous research, we avoid using accuracy as an evaluation metric due to the highly imbalanced nature of vulnerability fix detection datasets, where the majority class can dominate and lead to misleading results~\cite{he2009learning}. %F1-score and MCC provides a balance between precision and recall, offering metrics that considers both false positives and false negatives. Precision measures the accuracy of positive predictions, while recall measures the ability to identify all relevant instances. 

\subsection{RQ Approaches}
\subsubsection{\rqone}

In RQ1, we compare \ourTool with existing SOTAs, including three PLM-based techniques and three selected LLMs with different sizes under the CoT 0-shot setting.

First, we select three SOTA PLM-based techniques, including VulFixMiner~\cite{zhou2021finding}, CoLeFunda~\cite{zhou2023colefunda} and VulCurator~\cite{nguyen2022vulcurator}. We select VulFixMiner and CoLeFunda since they are the most commonly used baseline for vulnerability fix detection~\cite{pan2023fine,nguyen2023multi}.
VulCurator extends the approach proposed in VulFixMiner to include commit related development artifacts (\ie IR/PRs, commit message).
Due to its design, CoLeFunda's implementation is limited to the Java language, therefore, we only compare CoLeFunda with \ourTool on Java data.
We obtain these SOTA models as provided from the original authors, and use them according to guidance from the authors and or replication packages.
Note that we do not reuse the VulFixMiner dataset which was used to evaluate both VulFixMiner and CoLeFunda since this dataset poses a threat of data leakage since it contains data before 2023, which predates the cutoff dates of all LLMs included in our study. Therefore, we evaluate all models using our newly collected dataset BigVulFixes. 

Secondly, we also compare the LLMs under their vanilla setting without~\ourTool, defined as the 0-shot CoT setting.
For a fair comparison, we maintain the structure of the vulnerability fix detection prompt template as shown in~\Cref{box:vf_prompt}, but omit all information from the Code Change Intention (CCI), Development Artifact (DA), and History Vulnerability (HV) components. 
In other words, the LLM is only given the patch content and is tasked with determining whether the commit is a VF, without any additional context from~\ourTool. Additionally, the task instruction is simplified to: ``Determine whether the current patch is intended to fix a vulnerability. You must provide evidence if you think it's a vulnerability fix''.
% \sw{it is not very clear here, which templates are kept same, for example, without the three components, how would the template of vulnerablity fix detection work?}\xy{added detailed, please check}. 

\subsubsection{\rqtwo}
In our framework, we integrate three key components: Code Change Intention (CCI), Development Artifacts (DA), and Historical Vulnerability (HV). To understand the contribution of each component, we conduct an ablation study to evaluate their individual impact on the overall performance of \ourTool. 
For this study, we selected two models, Qwen2-72B and Qwen-7B, from the Qwen family. These models were chosen because both the larger and smaller versions demonstrated strong performance in our initial experiments (see section~\ref{sec:rq1_result} for details). In the ablation study, we systematically removed each component one at a time and compared the performance of the models with and without the removed component.

%Note that since not all commits are associated with development artifacts, the ablation study for the DA component was conducted only on data that included development artifacts, ensuring a fair comparison.

\subsubsection{\rqthree}
Having a model detect and label a commit as a VF is generally not the end of the story. 
Security experts will often perform screening on the instance to confirm the prediction was correct. This is an essential step before beginning remediation, such as applying the patch into downstream software dependencies. 
Therefore, in RQ3 we conduct a user study to investigate whether analysis generated by~\ourTool can help developers identify vulnerability fixes more effectively. We elaborate on the methodology of our user study below. 

\noindent\underline{Participants:} We invited 10 security experts from industry and academia with 3--5 years of software security experience for the user study from industry. We also ensure that these security experts experience of screening vulnerability fixes before. 

\noindent\underline{Task:} We randomly selected 40 VF and their analysis generated by~\ourTool based on Qwen2-72B.

\noindent\underline{Procedure:} Each participant is tasked to answer a sequence of yes or no questions with the option of adding additional comments given a VF and the generated analysis from~\ourTool. 

% \sw{I guess the purpose of this study, is to examine if developers have any difficulty to review predicted vul fix commit without any additional analysis information, only using raw data, right?}
% \sw{move this to online repository, if we need space... we are over the limit now}

We present the analysis result generated by \ourTool, and ask reviewers to answer the following questions:
\begin{enumerate}
    \item Does the analysis help understand the intent behind the code changes?
    \item Does the analysis accurately characterize the vulnerability?
    \item Does the analysis provide a better understanding of the root cause of the vulnerability?
    \item Does the analysis provided by the large model help improve the efficiency of identifying vulnerabilities fix?
    \item Are you satisfied with the quality of the generated content (\eg it contains redundant information, inaccurate information, not thorough enough, hallucination, historical vulnerabilities mentioned are irrelevant to the vulnerability, etc.)
\end{enumerate}

By collecting the answers to these questions, we aim to determine if the analysis information provided by \ourTool can help participants effectively understand the commit and verify vulnerabilities fixes.

\subsubsection{\rqfour} To better understand the limitations of \ourTool, we conduct a manual analysis on failed predictions, focusing on two types of misclassifications: (1) False Positives (FP): Cases where \ourTool incorrectly classifies a commit as a VF. This helps us examine scenarios where \ourTool misinterprets the intent or context of the commit. And (2) False Negatives (FN): Cases where \ourTool fails to identify a commit as a VF. This analysis highlights situations where \ourTool overlooks important indicators of a VF. 
% \begin{enumerate} 
% \item False Positives (FP): Cases where \ourTool incorrectly classifies a commit as a VF. This helps us examine scenarios where \ourTool misinterprets the intent or context of the commit. 
% \item False Negatives (FN): Cases where \ourTool fails to identify a commit as a VF. This analysis highlights situations where \ourTool overlooks important indicators of a VF. 
% \end{enumerate}

The first three authors each conducted a manual inspection on 20 FP and 20 FN for each of Qwen-72B and Qwen-7B following RQ2. A total of 240 cases has been reviewed.
We aim to identify the specific reasons for the misclassifications.
By pinpointing the root causes of the errors, we seek to understand the current limitations of \ourTool and identify room for future improvement.

\section{Results}
\subsection{RQ1 - Effectiveness}\label{sec:rq1_result}

\begin{table}[]
    \footnotesize
    \caption{The performance of vulnerability fix detection approaches.}\label{tab:rq1}
    \begin{tabular}{@{}lllcccc}
    \toprule
    \textbf{Foundation Model}                       & \textbf{Parameter Size} & \textbf{Approach}                               & \textbf{Precision}           & \textbf{Recall}              & \textbf{F1-score}                     & \textbf{MCC}                          \\ \midrule
                                        & 125M                    & VulFixMiner                                     & 0.17                         & 0.26                         & 0.20                                  & 0.14                                  \\
                                        & 125M                    & CoLeFunDa *                                     & 0.50                         & 0.06                         & 0.11                                  & 0.15                                  \\
    \multirow{-3}{*}{CodeBERT}          & 125M                    & VulCurator                                      & \textbf{0.77}                & 0.13                         & 0.22                                  & 0.30                                  \\ \midrule
                                        &                         & Vanilla                                         & 0.33                         & \textbf{0.84}                & 0.47                                  & 0.48                                  \\
                                        & \multirow{-2}{*}{236B} & \cellcolor[HTML]{EFEFEF}\ourTool & \cellcolor[HTML]{EFEFEF}0.40 & \cellcolor[HTML]{EFEFEF}0.78 & \cellcolor[HTML]{EFEFEF}0.53          & \cellcolor[HTML]{EFEFEF}\textbf{0.53} \\ \cline{2-7} 
                                        &                     & Vanilla                                         & 0.23                         & 0.44                         & 0.30                                  & 0.26                                  \\
    \multirow{-4}{*}{Deepseek-Coder-V2} &    \multirow{-2}{*}{16B}                     & \cellcolor[HTML]{EFEFEF}\ourTool & \cellcolor[HTML]{EFEFEF}0.28 & \cellcolor[HTML]{EFEFEF}0.65 & \cellcolor[HTML]{EFEFEF}0.39          & \cellcolor[HTML]{EFEFEF}0.37          \\ \midrule
                                        &                         & Vanilla                                         & 0.41                         & 0.53                         & 0.47                                  & 0.43                                  \\
                                        & \multirow{-2}{*}{70B}   & \cellcolor[HTML]{EFEFEF}\ourTool & \cellcolor[HTML]{EFEFEF}0.49 & \cellcolor[HTML]{EFEFEF}0.61 & \cellcolor[HTML]{EFEFEF}\textbf{0.54} & \cellcolor[HTML]{EFEFEF}0.52          \\ \cline{2-7} 
                                        &                         & Vanilla                                         & 0.10                         & 0.86                         & 0.18                                  & 0.18                                  \\
    \multirow{-4}{*}{Llama3.1}          & \multirow{-2}{*}{8B}    & \cellcolor[HTML]{EFEFEF}\ourTool & \cellcolor[HTML]{EFEFEF}0.30 & \cellcolor[HTML]{EFEFEF}0.50 & \cellcolor[HTML]{EFEFEF}0.37          & \cellcolor[HTML]{EFEFEF}0.34          \\ \midrule
                                        &                         & Vanilla                                         & 0.32                         & 0.72                         & 0.44                                  & 0.43                                  \\
                                        & \multirow{-2}{*}{72B}   & \cellcolor[HTML]{EFEFEF}\ourTool & \cellcolor[HTML]{EFEFEF}0.38 & \cellcolor[HTML]{EFEFEF}0.77 & \cellcolor[HTML]{EFEFEF}0.51          & \cellcolor[HTML]{EFEFEF}0.50          \\ \cline{2-7} 
                                        &                         & Vanilla                                         & 0.24                         & 0.48                         & 0.32                                  & 0.29                                  \\
    \multirow{-4}{*}{Qwen2}             & \multirow{-2}{*}{7B}    & \cellcolor[HTML]{EFEFEF}\ourTool & \cellcolor[HTML]{EFEFEF}0.52 & \cellcolor[HTML]{EFEFEF}0.48 & \cellcolor[HTML]{EFEFEF}0.50          & \cellcolor[HTML]{EFEFEF}0.47          \\ \bottomrule
    \end{tabular}\\
    * Due to the tool limitation, CoLeFunDa is evaluated on only Java vulnerabilities.
    \vspace{-0.15in}
    \end{table}

\subsubsection{PLM-based Approaches vs. \ourTool.}
%\jy{@xu @Cosmo, repharse with Logic: 1. LLM-based approach significantly outperformance PLM-based approach， especially the LLM4VFD solution. 2. For the recall, precision, F1-score, MCC, LLM4VFD outperform PLM-based models with any foundation LLM, in both small and large model sizes. 3. For the precision, Vulcurator and CoLeFunDa outperform LLM4VFD, however compromising the recall, making such high precision useless in practical scenario (missing the majority of VFs.)}

\sloppy\textbf{\ourTool outperforms PLM-based approach consistently in terms of MCC, F1-score, recall. For instance, \ourTool outperforms the best-performed PLM-based approach VulCurator by 68.1\% - 145.4\% across LLMs in terms of F1-score.}
As shown in Table~\ref{tab:rq1}, \ourTool significantly outperforms all three PLM-based approaches, VulFixMiner, CoLeFunda, and VulCurator, in terms of F1-score, MCC, and Recall.
\ourTool with various LLMs achieves an F1-score ranging from 0.37 to 0.54, which is significantly higher than PLM-based approaches with an F1-score of 0.14 to 0.22.
For instance, the best performed \ourTool (with Llama3.1-70B) achieves an F1-score of 0.54, exhibiting a 145.5\% improvement in F1-score compared to the best PLM-based approach VulCurator.
Although VulCurator and ColeFunDa achieve good precision 0.50 and 0.77 respectively, they suffer from very low Recall of 0.06 and 0.13 and miss many true vulnerability fixes, which hinders their practical application. 

\subsubsection{Vanilla LLMs (0-shot CoT) vs. \ourTool.}
%\sw{reorganize the text around the following key findings.}

\textbf{\ourTool outperforms all vanilla LLMs in terms of F1-score and MCC.}
Compared to the vanilla LLM, we observe that \ourTool improves performance across all evaluation metrics, except Recall on two LLMs (\ie Deepseek-Coder-V2 and Llama3.1-8B-Instruct).
In terms of F1-score, for each LLM \ourTool achieves a range of 12.7\% to 105.6\% improvements over its vanilla setting.
We observe a similar pattern of improvement for MCC.
In particular, \ourTool with Deepseekcoder-V2 achieves a Precision of 0.40 and a Recall of 0.78, resulting in an F1-score of 0.53, representing a 13\% improvement in Precision and a 12.8\% increase in F1-score compared to its vanilla setting.
Similarly, Llama3.1-70B improves its Precision from 0.41 to 0.49 (a 19.5\% increase) and Recall from 0.53 to 0.61 (a 15.1\% increase), resulting in an F1-score of 0.54 (a 14.9\% increase).

\textbf{Larger models typically outperform smaller models, while smaller models benefit more from our framework. For instance, F1-score for smaller models achieve an improvement of 64.0\% on average after applying \ourTool compared with the vanilla setting, while large models achieve an average improvement of 14.4\%.}
When comparing smaller models with their larger counterparts within the same family, we observed that the larger models consistently outperform the smaller ones, both with and without the \ourTool framework.
This is likely because vulnerability fix detection is a complex task that benefits from the extra parameters providing more code understanding capabilities.
In addition, the explanations for LLMs in the vanilla setting show significant variation, with F1-scores ranging widely from 0.18 to 0.50, especially for smaller models.
For example, within the Llama family, the vanilla Llama3.1-70B-Instruct model achieves an F1-score of 0.47, whereas the smaller Llama3.1-8B-Instruct model only achieves an F1-score of 0.18.
However, after applying our \ourTool framework, we find that the improvement in performance is more pronounced for smaller models than for larger ones.
On average, the F1-scores see a relative improvement of 64.0\%, whereas larger models saw a more modest average increase of 14.4\%.
These results indicate that \ourTool significantly enhances the performance of smaller models, making them more competitive with larger models and narrowing the performance gap.

%Notably, smaller models like Llama3.1-8B and Qwen2-7B benefit the most from \ourTool's enhancements. Llama3.1-8B sees its Precision rise from 0.10 to 0.30 and F1-score improve from 0.18 to 0.37, representing a 200\% improvement in Precision and 105\% improvement in F1-score. Similarly, Qwen2-7B's Precision improves from 0.24 to 0.51 and its F1-score rises from 0.32 to 0.49, reflecting a 112.5\% improvement in Precision and a 53\% improvement in F1-score. 

%The experimental results clearly demonstrate the superiority of \ourTool in enhancing the effectiveness of LLMs for vulnerability-fix detection. By integrating Code Change Intention, Historical Vulnerability Data, and Development Artifacts, \ourTool provides a more comprehensive understanding of commits, leading to notable improvements in Precision and Recall across all LLMs. This design enables the models to detect vulnerabilities more accurately than traditional PLM-based techniques. The significant performance gap between \ourTool and vanilla LLMs highlights the limitations of vanilla LLM approaches in capturing the complex semantics of code changes. In contrast, \ourTool’s structured approach, which leverages Chain-of-Thought (CoT) and Retrieval-Augmented Generation (RAG), significantly enhances the LLM's capacity to reason about code changes, resulting in higher performance.

\rqboxc{\ourTool outperforms PLM-based approach consistently in terms of MCC, F1-score, and recall. For instance, \ourTool outperforms the best-performed PLM-based approach VulCurator by 68.1\%--145.4\% across LLMs in terms of F1-score. Our framework demonstrates performance gains ranging from 12.7\% to 105.6\% over its vanilla variant, with smaller models generally benefiting more compared to their larger counterparts.}

\subsection{RQ2 - Ablation analysis.}\label{sec:ablation}

\begin{table}[]
\caption{The ablation results. The cells with the lowest performance are marked in bold, indicating the largest contribution from that component.}\label{tab:rq2}
\footnotesize
\begin{tabular}{@{}lcccclcccc@{}}

\toprule
                          \textbf{Ablation Setting}& \multicolumn{4}{c}{\textbf{Qwen2-72B}}                         &  & \multicolumn{4}{c}{\textbf{Qwen2-7B}}                          \\ \cmidrule(r){2-5} \cmidrule(l){7-10}
                          & \textbf{Precision} & \textbf{Recall} & \textbf{F1-score} & \textbf{MCC} &  & \textbf{Precision} & \textbf{Recall} & \textbf{F1-score} & \textbf{MCC} \\ \cmidrule(r){1-5} \cmidrule(l){7-10} 
\ourTool   & 0.38               & 0.77            & 0.51              & 0.50         &  & 0.52               & 0.48            & 0.50              & 0.47         \\ 
\rule{0pt}{2.5ex}\hspace{0.2cm}w/o CCI component & \textbf{0.33}               & 0.77            & \textbf{0.46}              & \textbf{0.45}         &  & 0.40               & 0.54            & \textbf{0.46 }             & \textbf{0.42}         \\ 
% w development artifact    & 0.35               & 0.76            & 0.48              & 0.48         &  & 0.54               & 0.50            & 0.52              & 0.50         \\
\hspace{0.2cm}w/o DA component  & 0.36               & 0.75            & 0.48              & 0.48         &  & 0.49               & \textbf{0.47}            & 0.48              & 0.45         \\ 
% w history vulnerability   & 0.38               & 0.77            & 0.51              & 0.50         &  & 0.51               & 0.48            & 0.50              & 0.47         \\
\hspace{0.2cm}w/o HV component & 0.36               & \textbf{0.74}            & 0.49              & 0.48         &  & \textbf{0.39 }              & 0.62            & 0.48              & 0.45         \\ \bottomrule
\end{tabular}
\end{table}

\textbf{Overall, all three components in \ourTool make positive contributions to the overall performance. Among them, Code Change Intention have a larger impact than Development Artifact and Historical Vulnerability.} Table~\ref{tab:rq2} shows the results of our ablation analysis. 
The removal of CCI component leads to a significant decrease in performance, particularly in terms of Precision and F1-score. For Qwen2-72B, the Precision drops from 0.38 to 0.33, representing a 13.1\% reduction. In Qwen2-7B, the decline is similar, with the Precision decreasing from 0.52 to 0.40, an 15.4\% reduction. These results emphasize the important role of CCI in improving the model's ability to capture vulnerability fix commits more accurately by reducing false positives, and the improvement is especially notable in smaller models, where CCI helps balance Precision and Recall effectively. DA component has a more similar effect on Qwen2-7B and Qwen2-72B. In Qwen2-72B, the F1-score improves significantly from 0.48 to 0.51, marking an 6.3\% increase, indicating that the additional contextual information provided by DA substantially boosts both precision and recall.
% In contrast, Qwen2-72B shows only a slight improvement in F1-score from 0.46 to 0.48, suggesting that larger models may be less dependent on external development context, while smaller models see considerable gains from DA. 
% Removing the HV component results in slight reductions in Precision and F1-score for both models. In Qwen2-72B, the F1-score drops from 0.51 to 0.49 with a 3.9\% decrease, while Qwen2-7B sees a more significant reduction in Precision from 0.51 to 0.39, a 23.5\% decrease. HV's contribution is primarily observed in its ability to reduce false positives by grounding predictions in historical vulnerability data, thereby improving Precision without significantly impacting Recall.
Removing the HV component or DA components results in slight reductions in F1-score and MCC for both models. However, we find that Qwen2-7B sees a more significant reduction in precision without the HV components, dropping from 0.52 to 0.39, a 25\% decrease. 

The ablation study highlights the importance of each component in \ourTool, particularly CCI, which demonstrates significant improvements in precision and F1-score, especially for smaller models like Qwen2-7B. HV further enhances precision by reducing false positives, albeit with a more moderate effect. Collectively, these components provide a comprehensive framework that significantly enhances the accuracy and reliability of vulnerability fix detection, particularly by augmenting smaller models with crucial contextual and historical information.
% \mike{should we briefly mention any reasons that HV component might contribute less to the overall performance (e.g., distance function can be improved, embedding and query model can be improved, indexing can be improved like chunking, ...}

% \xy{need to rewrite RQ result based on new result}
\rqboxc{All components in \ourTool make positive contributions to the performance. The impact from Code Change Intention is larger than Development Artifact and Historical Vulnerability.}

\subsection{RQ3 - User study}
\label{sec:rq3}

\textbf{The analysis generated by \ourTool helps security experts to understand the intent of code changes and the vulnerabilities, which improves the efficiency of identifying vulnerability fixes.}
% Before being shown the analysis generated by \ourTool, participants found that they could not understand the purpose of the code changes in 27.5\% of the commits (Question 1), while this percentage reduced to 5\% after the participants were provided the LLM-generated analysis (Question 3)\sw{I dont' think question 3 can reflect this. one option we only keep the second part of the question.}\mike{I agree this claim seems odd to me / Q3 does not really reflect this. I notice here in text you mention ``before seeing the analysis'' but what you wrote as Q1 in section 5 does not reflect this so it is confusing. If they truly did not see it before Q1, then Q1 is applicable here, but still dont see the connection to Q3. So we can do what Shaowei suggests, or maybe you can explain it in another way / how Q3 shows reflects this claim}. 
Our result shows that in 95.0\% (38 out of 40) of the cases, \ourTool’s analysis can help with understanding the intent of commits (Question 1). A case like CVE-2024-29199~\cite{nvd.2024.29199} has 47 changed files with 517 lines of addition and 226 lines of deletion. It is too long for participants to understand, however \ourTool's analysis helped participants to understand the change.
In addition, participants think that \ourTool accurately describes the characteristics of the vulnerabilities in 90\% of the commits (Question 2) and the root causes in 75.0\% (30 out of 40) of the cases (Question 3).
With \ourTool, participants can identify VF commit more efficiently.
This is evidenced by the fact that in 80.0\% (32 out of 40) of the commits, \ourTool improved their efficiency in identifying vulnerability fixes (Question 4).
We investigate the 8 cases where participants did not think the analysis generated by \ourTool helped them to identify vulnerability fixes more efficiently, 7 of them are due to the vulnerability fixes are easy to identify even without the help of \ourTool.
In one case of CVE-2024-28103~\cite{nvd.2024.28103}, the participant did not understand the vulnerability even with the help of \ourTool.

% Unable to address ATM
%\mike{Is it worth it to have a metric to quantify the improvement in efficiency in another way? I am thinking about this but it seems difficult at this point. Maybe we can make a stronger claim, saying something like ``according to participants our user study shows our tool improves efficiency in 80\% of cases''}

%\ourTool's analysis was able to help Participants truly understand the root cause of the vulnerability. In CVE-2024-24762~\footnote{\url{https://nvd.nist.gov/vuln/detail/CVE-2024-24762}}, \ourTool failed to recognize the root cause, the analysis mentioned the vulnerability as a general ``security vulnerability'' instead of the specific ``DoS'' vulnerability. The result suggests that future research should put more effort into mining the root cause of vulnerability.

When investigating the feedback we collected related to the quality of the analysis generated by \ourTool (Question 5), we find that in 75\% of the cases, participants are satisfied with the overall quality of the analysis.
There were some cases (\eg CVE-2023-48014~\cite{nvd.2023.48014} and CVE-2023-37061~\cite{nvd.2023.37061}) where participants noted that the analysis contained redundant information, or lacked depth analysis for more complex vulnerabilities (\eg CVE-2023-48657~\cite{nvd.2023.48657}).
This suggests a need for further refinement in handling edge cases and ensuring that the historical context provided is directly relevant and concise. Nevertheless, ours is the first work on this direction and more future research is encouraged. 

%Overall, the study shows that \ourTool significantly enhances security experts’ ability to understand and efficiently identify VF commits.
% complex cases where the commit message or code diff alone is insufficient.

\rqboxc{Overall, the user study shows that the analysis generated by \ourTool helps security experts to understand the intent of code changes and the vulnerabilities, which improves the efficiency of identifying vulnerability fixes.}
\subsection{RQ4 - Failure analysis}

\begin{table}[]
\footnotesize
\centering
\caption{Bad Case Analysis on Qwen2 Models}\label{tab:rq4}
\begin{tabular}{@{}clrr@{}}
  \toprule
% \multicolumn{3}{|c|}{\textbf{Qwen2-72B-Instruct}} & \multicolumn{3}{c|}{\textbf{Qwen2-7B-Instruct}} \\ \hline
\textbf{Type}                                     & \textbf{Reason}                                                        & \textbf{72B} & \textbf{7B} \\ \midrule
\multirow{6}{*}{FP}                               
                                                  & Potential unreported vulnerability fix                                 & 1              & 7 \\
                                                  & Non-vulnerability security fix misclassified as vulnerability fix                        & 45             & 36 \\
                                                  & Non-functional Change                                                  & 7              & 8 \\
                                                  & Fail to realize the change is not related to security                  & 4              & 5 \\
                                                  & Mislead by retrieved similar vulnerability                             & 2              & 3 \\
                                                  & Others                                                                 & 1              & 1 \\ \midrule
\multirow{5}{*}{FN}                               & Vulnerability fix misclassified as non-vulnerability security fix                        & 28             & 33 \\
                                                  & Unable to identify security related code change                        & 21             & 18 \\
                                                  & Mislead by retrieved similar vulnerability                             & 5              & 6 \\
                                                  & Unable to pinpoint vulnerability related code change from long context & 1              & 0 \\
                                                  & Others                                                                 & 5              & 3 \\ \bottomrule
\end{tabular}

\end{table}
%\xy{say clear relation between sec fix and vuln fix, some example of sec fix not vuln fix}\xy{emph explanation is useful}

% \xy{add two sentence overview}
% \sw{what is the difference between security fix and vulnerability fix? }\xy{vulnerability fix is a subset of security fix}

Table~\ref{tab:rq4} summarizes the results of our failure analysis. To better understand the misclassification issues encountered by \ourTool, it is crucial to clarify the relationship between security fixes and vulnerability fixes. A \emph{vulnerability fix} is a subset of \emph{security fixes}. Security fixes encompass any code changes that enhance the overall security of the software, such as improvements to authentication mechanisms, encryption implementations, or adherence to security best practices, whereas vulnerability fixes aim to address security flaw, glitch, or weakness found in software code that could be exploited by an attacker~\cite{nvd.homepage}. Therefore, in our dataset, a security fix commit that does not aim to fix a vulnerability is considered as non-vulnerability fix (i.e., \textit{non-vulnerability security fix}), although they are related to security fix.

The most mis-classification made by \ourTool due to its struggling to distinguish between vulnerability fix and a non-vulnerability security fix. 
For instance, in false positive (FP) cases, 75.0\% of FP in Qwen2-72B and 60.0\% in Qwen2-7B occured due to non-vulnerability security fixes are misclassified as vulnerability fixes. 
Vise versa, it led to 46.7\% and 55.0\% of the FN cases where vulnerability fixes are misclassified as non-vulnerability fixes in Qwen2-72B and Qwen2-7B. The result indicates that even if \ourTool can identify security-related commits, it sometimes fails to interpret their severity. Specifically, in the analysis of an FN case, the LLM explains with: ``Although changes are related to security, I am uncertain if the changes fix a vulnerability''.
One possible explanation is that the LLM is unable to differentiable between security related change and vulnerability fix when generating the intention for commits.
Although, \ourTool struggles to distinguish between vulnerability fix and non-vulnerability security fix. Both vulnerability fix and non-vulnerability security fix are security fixes, and identifying them is beneficial. 

Another frequent failure in FN cases is the inability to identify security related code change (32.5\%). Another notable failure case is related to its use of historical vulnerability data through the HV component. The retrieval of irrelevant history vulnerability causes misclassification in both FP and FN cases. Specifically, in 4.2\% of the FP cases, and 9.2\% of the FN cases, the HV component retrieved vulnerabilities that seemed relevant but did not match the functional or contextual details of the current commit, and such information caused confusion to LLM, leading \ourTool to make incorrect assumptions on the commits. While our HV component improves the overall performance (as shown in our ablation study in \Cref{sec:ablation}), it is not perfect and can sometimes retrieve irrelevant historical vulnerabilities. Future studies can focus on this aspect by developing better retrieval techniques and improving the underlying vulnerability database. 

We also find in a few FP cases where the commit is potentially fixing an unreported vulnerability.
For example, in one of the checked cases, the commit is a merge of changes from a GHSA~\cite{github.ghsa} pull request.
GHSA is an independent security advisor that maintains its own vulnerability dataset and may have different assessments compared to NVD.

%\xy{shaowei, can we say this? seems to attack our approach}The bad cases above indicate that the current level of context provided to LLM by \ourTool is not sufficient for LLM to clearly identify vulnerability fix from security fix. However, the detailed analysis and explanations generated by \ourTool can help user  In future work, we will further improve context engineering, provide more relevant context and mine deeper information.

% \textbf{Mislead by retrieved similar vulnerability}
%     % \item \textbf{Insufficient Context Information.} Insufficient context information also cause 
%     % \item \textbf{}

% In addition, the reasons for false negativew can be summarized into the following categories:
% \begin{itemize}
%     \item Unable to identify security related code change
% \end{itemize}

\section{Discussion}

% \subsection{Compromising between Precision and Recall}
% PLM approaches typically output a probability to indicate the likelihood of a commit being made for a vulnerability fix. A threshold between 0 and 1 is provided to make final binary decision.
% By adjusting the threshold value, we could favor precision or recall based on the needs.
% However, this technique is no longer applicable when using LLM since LLM's strength is in generating natural language outputs.
% During our prompt engineering phase, we noticed that by tweaking the word choice in the system and user prompt, we can instruct the LLM to favor precision or recall in the final prediction.
% However, this adjustment is more subtle and hard to quantify. We encourage future work to further explore this aspect and design LLM-based systems that allow some flexibility in favoring precision or recall.

\subsection{Potential Future Direction}
Our research is the first study to explore vulnerability fix detection using LLMs. However, insights from our user study and failure case analysis indicate several areas for future exploration. First, incorporating more relevant project-specific information, such as security policies and historical commit patterns, could help LLMs better distinguish vulnerability fixes among security-related commits. Second, the prompt templates guiding the LLM could be refined to more effectively leverage context and produce more insightful analysis results for security experts. Finally, the current RAG design is basic, and incorporating advanced techniques --- such as dynamic retrieval strategies or reranking mechanisms --- could improve the precision and relevance of retrieved historical vulnerabilities.

\subsection{Threats to Validity}

\subsubsection{Internal Validity}
Due to the extremely imbalanced nature of vulnerability fixes and non-vulnerability fixes, when our approach is applied to monitor real-world projects there may be a large amount of false negatives. Based on feedback from our industry partners, the false positive rate was considered acceptable, as it requires only a small amount of additional human effort to review the results. Previous studies suggest that LLM settings, such as temperature, have an impact on outputs~\cite{renze2024effect,peeperkorn2024temperature}. In this study, we use the default settings for the studied LLMs for all RQs.

\subsubsection{External Validity}
Threats to external validity relate to the generalizability of our approach. Since \ourTool is a framework, and its implementation relies on LLMs. The effectiveness of \ourTool also depends on the performance and capabilities of the LLMs. To mitigate this threat, we evaluated \ourTool using several well-known and state-of-the-art LLMs, including both large models (70B--236B parameters) and smaller models (7B--16B parameters), and our results show that with all models, \ourTool outperforms SOTA baselines. Future research is encouraged to investigate the performance with more LLMs using our framework.

\section{Conclusion}
In this paper, we propose \ourTool, a novel framework that leverages Large Language Models (LLMs) enhanced with Chain-of-Thought reasoning and In-Context Learning to improve the accuracy of vulnerability fix detection by integrating information from multiple sources (e.g., related development artifacts and historical vulnerabilities). More importantly, on top of the prediction, \ourTool also provides a detailed explanation in analysis to help security experts understand the rationale behind the decision.
Experimental results demonstrate that \ourTool outperforms PLM-based approach consistently in terms of MCC, F1-score, and recall. For instance, \ourTool outperforms the best-performed PLM-based approach VulCurator by 68.1\% - 145.4\% in terms of F1-score when using different LLMs as the base model.
Furthermore, we conducted a user study involving security experts to assess the effectiveness of the analysis generated by \ourTool in aiding vulnerability fix identification. The feedback from the participants demonstrates that the analysis provided by \ourTool improves the efficiency of identifying vulnerability fixes.

\section*{Data Availability}
We make all the datasets and code used in this study openly available in our replication package~\cite{anonymous_2024_13776994}.

%\section*{Acknowledgment}

\bibliographystyle{ACM-Reference-Format}
\bibliography{references}

\end{document}